\documentclass{Interspeech}

\interspeechcameraready

\title{Fairness in Dysarthric Speech Synthesis: Understanding Intrinsic Bias in Dysarthric Speech Cloning using F5-TTS}

\author[affiliation={1}]{Anuprabha}{M}
\author[affiliation={2}]{Krishna}{Gurugubelli}
\author[affiliation={1}]{Anil Kumar}{Vuppala}

\affiliation{Speech Processing Lab, LTRC}{International Institute of Information Technology-Hyderabad}{India}
\affiliation{}{Samsung Research \& Development Institute-Bengaluru}{India}
\email{anuprabha.m@research.iiit.ac.in, krishna.g@samsung.com, anil.vuppala@iiit.ac.in}

\newcommand{\blue}[1]{\textcolor{blue}{#1}}
\usepackage{comment}
\usepackage{algorithm2e}
\usepackage{subfigure}
\usepackage{multirow}

\begin{document}

\maketitle

\begin{abstract}
Dysarthric speech poses significant challenges in developing assistive technologies, primarily due to the limited availability of data. Recent advances in neural speech synthesis, especially zero-shot voice cloning, facilitate synthetic speech generation for data augmentation; however, they may introduce biases towards dysarthric speech. In this paper, we investigate the effectiveness of state-of-the-art F5-TTS in cloning dysarthric speech using TORGO dataset, focusing on intelligibility, speaker similarity, and prosody preservation. We also analyze potential biases using fairness metrics like Disparate Impact and Parity Difference to assess disparities across dysarthric severity levels. Results show that F5-TTS exhibits a strong bias toward speech intelligibility over speaker and prosody preservation in dysarthric speech synthesis. Insights from this study can help integrate fairness-aware dysarthric speech synthesis, fostering the advancement of more inclusive speech technologies.

\end{abstract}

\section{Introduction}

Dysarthria, a motor speech disorder caused by neurological conditions, results in a speech that is often slurred, slow, or difficult to understand, particularly in its more severe forms~\cite{joseph2019motor, enderby2013disorders}. Individuals with dysarthria often encounter communication challenges, making speech-based assistive technologies such as Automatic Speech Recognition (ASR) and Text-to-Speech (TTS) systems vital for facilitating their interaction with digital devices. However, the development of robust dysarthric speech technologies is often hindered by data scarcity, as dysarthric speech varies widely in severity and speaker characteristics~\cite{rowe2022characterizing}. 
Data augmentation is widely explored in dysarthric speech applications to address this challenge of data scarcity~\cite{geng20_interspeech}. 

\hbadness=10000
Traditional approaches for data augmentation include speed, temporal, or spectral modifications~\cite{geng20_interspeech, jin2024towards,bhat22_interspeech}, GAN-based adversarial augmentation~\cite{jin2023personalized,wang2024enhancing}, and voice conversion~\cite{harvill2021synthesis, huang2022towards,halpern2023improving}.
Additionally, advancements in end-to-end TTS and voice cloning systems have enabled the generation of high-quality synthetic speech with human-like qualities, including diverse speaker characteristics and styles. These advancements have been studied by~\cite{azizah2024zero, hermann23_interspeech, hu2023generating} to synthesize dysarthric speech.
By leveraging state-of-the-art voice cloning models, such as F5-TTS~\cite{chen2024f5}, researchers can generate personalised dysarthric speech at varying severity levels, which has the potential to enhance the training of ASR models and other assistive speech technologies. Despite the utilization of neural speech synthesis systems for augmenting dysarthric speech, several critical questions remain unanswered regarding the effectiveness and fairness of dysarthric speech cloning, like: 
\textit{Can synthetic dysarthric speech maintain intelligibility? How does it impact ASR performance? How well the cloned speech retains speaker identity in replicating different severity levels of dysarthria? Do speech synthesis models exhibit bias across dysarthric severity levels in ways that could be disadvantageous to certain user groups?}

During dysarthric speech synthesis across severity levels, challenges related to fairness can manifest in various ways, such as intelligibility bias (comprehensibility of spoken content), speaker identity (preservation of speaker characteristics), and prosody inconsistencies (rhythmic and tonal aspects of speech). 
Due to this, speech data augmentation becomes biased, leading to skewed performance in ASR and making speech recognition less effective, particularly for severely impaired dysarthric speakers~\cite{zhang2023exploring}. Moreover, assistive speech systems trained on biased synthetic data could reinforce disparities, thus negatively impacting accessibility and usability for individuals with dysarthria. Hence, it is very important to understand the intrinsic biases in dysarthric speech syntheses to incorporate fairness-aware augmentation strategies. 
Over the past decade, significant work has been done in assessing biases by focusing on age, gender, and geographic factors within ASR and spoken language understanding (SLU) systems~\cite{dheram22_interspeech, lai23_interspeech, veliche24_interspeech, koudounas2024towards}. To the best of our knowledge, this is the first study to analyse biases in synthetic dysarthric speech generated by a zero-shot voice cloning system.  The key contributions of this work are as follows:

\begin{itemize}
    
    \item Assess the quality of synthesised dysarthric speech in terms of intelligibility, speaker similarity, and prosody preservation, using objective metrics like Word Error Rate (WER), Character Error Rate (CER), Speaker Similarity (SIM-o) and Auto Prosodic Consistency Protocol (AutoPCP).

    \item Introduce a framework to quantify fairness in synthesising dysarthric speech based on Parity Difference (PD) and Disparate Impact (DI) metrics. 

    \item An in-depth fairness analysis is conducted on dysarthric speech generated by F5-TTS, examining variations across different severity levels, gender, and individual speakers using the TORGO dataset~\cite{TORGO}. 

    \item Impact of incorporating voice-cloned samples on downstream tasks, such as detection and speech recognition of dysarthria.
    \item Provision of key insights into the need for fairness-aware analysis in the generation of synthetic data to improve dysarthric speech technologies, particularly ASR and assistive communication systems.
        
\end{itemize}

\section{Proposed Methodology}
This section introduces the proposed framework for evaluating performance variations in preserving information related to dysarthric severity during zero-shot synthetic speech generation. As illustrated in Figure~1, the framework follows a systematic approach, beginning with the generation of synthetic speech using F5-TTS, followed by an objective assessment, and concluding with the measurement of bias across different severity categories using two fairness metrics. Furthermore, this section outlines the formulation and methodology employed to derive the fairness metrics based on the objective assessment.

\begin{figure}[htbp!]
      \centering
      \vspace{-5pt}
      \includegraphics[width=\linewidth]
      {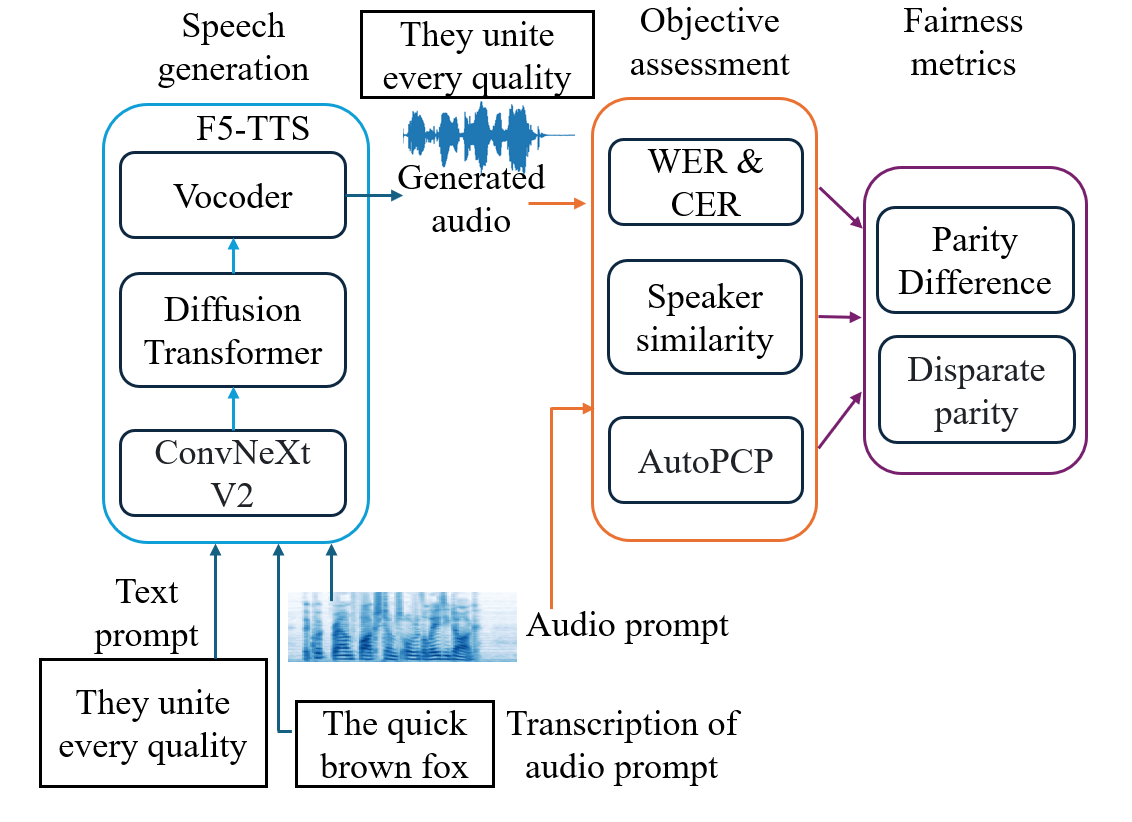}
      \caption {Framework proposed for assessing fairness in dysarthric speech generation using F5-TTS}

\end{figure}
\vspace{-6mm}

\subsection{Speech Cloning using F5-TTS}

F5-TTS~\cite{chen2024f5}, a zero-shot, non-autoregressive TTS system, is employed in this study to generate speech samples for dysarthric speakers. The model synthesizes speech using a few seconds of an audio prompt (reference speech sample), its correspoding transcription, and a text prompt for synthetic speech generation. F5-TTS leverages flow matching and the Diffusion Transformer (DiT), eliminating the need for components such as a duration predictor, text encoder, context-aware codec model, or phoneme alignment. Instead, it employs ConvNeXt V2 to align the text input with the speech length by padding it with filler tokens, followed by a de-noising process to generate high-quality, natural-sounding speech. Trained on 100K hours of publicly available multilingual data~\cite{he2024emilia}, F5-TTS exhibits exceptional zero-shot performance, producing realistic speech while preserving naturalness, intelligibility, and speaker similarity. In this study, the F5-TTS base model~\footnote{\url{https://github.com/SWivid/F5-TTS}} is used to generate synthetic speech for speakers in the TORGO database.

\subsection{Objective Assessment of Dysarthric Speech Cloning}
To objectively assess how well the generated speech preserves different levels of dysarthric severity in terms of intelligibility, speaker, and prosodic similarity, we evaluate it using the following criteria:
\begin{itemize}
\item \textbf{WER} and \textbf{CER}:
The intelligibility of the generated samples is assessed using both WER and CER. The addition of CER offers a detailed analysis by focusing on errors at the character level, which is particularly useful for capturing subtle speech distortions such as mispronunciations or missing characters often present in dysarthric speech~\cite{choi2024speech}. Due to this, the evaluation process gains a more comprehensive understanding of the TTS system's ability to handle the intricate nuances of dysarthric speech, ensuring a more accurate assessment of its overall performance on intelligibility.
Additionally, to evaluate potential biases in the context of intelligibility, it is essential to incorporate the differences ($\Delta \text{WER}$) between WER of the audio prompt and the generated audio as part of the objective assessment. $\Delta \text{WER}$ is obtained by:
\begin{align}
\Delta \text{WER} = WER_\text{audio prompt} - WER_\text{generated audio}
\end{align}
\vspace{-2pt}
Similarly,  $\Delta \text{CER}$ is calculated by,
\vspace{-2pt}
\begin{align}
\Delta \text{CER} = CER_\text{audio prompt} - CER_\text{generated audio}
\end{align}

In this study, the Hubert-large\footnote{\url{https://github.com/facebookresearch/fairseq/tree/main/examples/hubert}} ASR model~\cite{hsu2021hubert} fine-tuned on 960~hrs of Librispeech is used to measure WER and CER.
\item \textbf{SIM-o score}:
Assessing speaker similarity is essential to ensure that the generated speech accurately reflects the intended speaker's characteristics despite potential distortions. In this study, speaker similarity is evaluated using the SIM-o score, which quantifies the preservation of speaker characteristics. SIM-o is computed as the cosine similarity between the speaker embeddings of the reference audio prompt and the generated speech. To obtain SIM-o, we utilize a speaker verification model\footnote{\url{https://github.com/microsoft/UniSpeech/tree/main/downstreams/speaker_verification}} based on WavLM-large~\cite{chen2022wavlm}. 

\item \textbf{AutoPCP score}~\cite{barrault2023seamless}:
For dysarthric speakers, maintaining natural prosody is challenging due to motor impairments that affect speech articulation, pitch, rhythm, and intonation, often resulting in irregular speech patterns. Evaluating prosody similarity is crucial to ensure that the generated speech preserves the prosody of the original audio prompt. This is calculated by comparing the audio prompt with the generated speech using AutoPCP\_multilingual\_v2
\footnote{\url{https://github.com/facebookresearch/stopes/tree/main/stopes/eval/auto_pcp}} comparator model.
\end{itemize}

\subsection{Fairness Metrics}
\begin{itemize}
\item Parity Difference (PD):
It refers to the extent of similarity or differences in the objective measures between healthy and various severity categories. PD value of 0 indicates that the categories are treated similarly.
\item Disparate Impact (DI):
The Disparate Impact, or relative disparity, is the ratio calculated for each of the objective measures pertaining to healthy and various severity categories. DI value of 1 indicates no bias between the groups. 

\end{itemize}
\subsection{Formulation / Algorithm}

\begin{algorithm}
\caption{Estimation of fairness metrics.}
\While{$d \in D$}{
    \While{$s \in S$}{
        $ PD_{d,s} = | m_{d,s=\text{healthy}} - m_{d,s} | $;
    }    
  \eIf{$d$ is $\Delta \text{WER}$ or $\Delta \text{CER}$}{
        $ m_{d,s} = Softmax(m_{d,s}) $ \;
        \While{$s \in S$}{
        $ DI_{d,s} = \frac{1/m_{d,s}}{1/m_{d,s=\text{healthy}}}$;
        }
  }{    \While{$d \in D$}{
        $ DI_{d,s} = \frac{m_{d,s}}{m_{d,s=\text{healthy}}}$;
        }
  }
}

\end{algorithm}

Let $s \in \{ healthy, low, mid, high \} $ denote a severity category from the set of $S$ severities listed in Table 1 and $d \in \{ \Delta WER, \Delta CER, SIMo, AutoPCP \} $ denote an objective measure from a set of $D$ objective measures. 
$m_{d,s}$ is the mean of the objective measure $d$ for $s$ severity category. Based on the  information provided, the algorithm for measuring bias across different categories using fairness metrics is proposed and outlined in Algorithm~1. While using $\Delta \text{WER}$ or $\Delta \text{CER}$ to measure bias, higher word (or character) error rates may result in higher severity categories, potentially skewing the results. To address this, softmax normalization is applied to the mean of $\Delta \text{WER}$ or $\Delta \text{CER}$ across the severity levels to effectively find a more balanced DI. Softmax normalization for $d$ is given by:
\begin{align}
\text{Softmax}(m_{d,s}) = \frac{e^{m_{d,s}}}{\sum_{S} e^{m_{d,s}}}
\end{align}

\section{Experimental setup}
This section provides an overview of the database and outlines the evaluation setup for the study.
\vspace{-0.1cm}
\subsection{TORGO Dysarthric Database}
The TORGO database~\cite{TORGO} includes both acoustic and 3D articulatory measurements from speakers with Cerebral Palsy (CP) or Amyotrophic Lateral Sclerosis (ALS). It contains speech samples from 8 dysarthric (3 females and 5 males) and 7 healthy (3 females and 4 males) speakers. The utterances in the database are categorized into: non-words, short words,  restricted sentences, and unrestricted sentences. Utterances from restricted sentences are considered for this study. The dysarthric speakers in the database were assessed using the Frenchay Dysarthria Assessment (FDA), a 9-point scale that measures speech motor function and severity. Table~\ref{table:torgo} presents the list of speakers with varying severity levels included in this study. 
\vspace{-8pt}
\begin{table}[h!]
\centering
\caption{Speaker Severity Categories in the TORGO database. All of these speakers are considered for generating synthetic samples, and train-test split is followed in the downstream tasks.}
\label{table:torgo}
\vspace{-9pt}
\begin{tabular}{ccc}
\toprule
\textbf{Severity} & \textbf{Train} & \textbf{Test} \\ \midrule
healthy           & FC03, MC03     & MC04          \\
low               & F04, M03       & F03           \\
mid               & M05            & F01           \\
high              & M01, M04       & M02           \\ \bottomrule
\end{tabular}
\vspace{-15pt}
\end{table}

\subsection{Downstream Tasks Training and Evaluation}
To perform the fairness analysis, for each speaker listed in Table~\ref{table:torgo}, $n$ reference audio prompts are considered, and $3n$ synthetic samples are generated using F5-TTS for given text prompts~\footnote{\url{https://github.com/microsoft/e2tts-test-suite/blob/main/assets/e2tts_librispeech_pc_test_clean.json}}.
For fine-tuning on the ASR task, Wav2vec2.0-Base~\cite{baevski2020wav2vec} model~\footnote{\url{https://github.com/facebookresearch/fairseq/blob/main/examples/wav2vec/README.md}} is employed without a language model (LM).
To perform dysarthria detection (healthy Vs. dysarthria), 39-dimensional MFCC features are extracted from speech for every 25~ms with a shift of 10~ms.
A CNN architecture with 3 convolutional layers, flattening and fully connected layers was utilized and each layer is provided with batch normalization, a Relu activation function, and a dropout of 0.2, to prevent overfitting. The downstream tasks were performed on data containing (1) only audio samples from the database; (2) only on generated audio; and (3) combination of both (1) and (2).
We used two NVIDIA GeForce RTX 2080 Ti GPUs.

\begin{table*}[t!]
\centering
\small\addtolength{\tabcolsep}{-4.5pt}
\caption{Key insights and recommendations based on the fairness study. (Based on minimum DI score of each objective measure.)}
\label{table:key_insights}
\begin{tabular}{|c|c|c|c|c|c|}
\hline
\textbf{Metric}  & \textbf{Most Affected Severity} & \textbf{DI Score} & \textbf{Fairness Level} & \multicolumn{1}{c|}{\textbf{Key Insights}}                                                                     & \multicolumn{1}{c|}{\textbf{Recommended Action}}                                                         \\ \hline
\textbf{$\Delta$WER}     & Mid and high severity         & 0.59              & Poor                    & \begin{tabular}[c]{@{}c@{}} High disparity in speech \\ recognition for severe dysarthria.\end{tabular}             & \begin{tabular}[c]{@{}l@{}}Intelligibility-aware speech data \\ augmentation is needed.\end{tabular}                     \\ \hline
\textbf{$\Delta$CER}     & Mid and high severity          & 0.75              & Poor                    & \begin{tabular}[c]{@{}c@{}}Character-level errors increase \\ notably for severe dysarthria.\end{tabular} & \begin{tabular}[c]{@{}l@{}}Adapt ASR models to represent   \\ severity for better generalisation.\end{tabular} \\ \hline
\textbf{SIM-o}   & No major impact                 & 0.81              & Good                    & \begin{tabular}[c]{@{}c@{}}Speaker identity is mostly \\ preserved across severity levels.\end{tabular}         & Minimal intervention is needed.                                                                              \\ \hline
\textbf{AutoPCP} & No major impact                 & 0.87              & Good                    & \begin{tabular}[c]{@{}c@{}}Prosody preservation remain \\ stable across dysarthria levels.\end{tabular}            & Minimal intervention is needed.                                                                              \\ \hline
\end{tabular}
\vspace{-10pt}
\end{table*}

\section{Results and Discussion}
This section provides findings and insights into intrinsic biases in dysarthric speech cloning using F5-TTS, highlighting their impact on fairness in dysarthric speech synthesis.
\vspace{-0.1cm}
\subsection{Objective Assessment of Synthetic Dysarthric Speech}
To assess how well the generated speech preserves dysarthric severity, different objective measures (as discussed in Section~2.2) are computed between the reference audios from TORGO dataset and generated audios using F5-TTS, and the results are presented in Figure~2. From Figure~2.a., it is observed that the $\Delta \text{WER}$ and $\Delta \text{CER}$ are consistently higher in mid and high severity categories compared to the healthy and low severity category. The main reason for this bias might stem from the limitations of the zero-shot voice-cloned sample in accurately representing the speech intelligibility of mid and high severity categories. This results in a lower WER or CER in the generated audio compared to the reference. 

In Figure 2(b), the SIM-o and autoPCP metrics assess speaker similarity and prosody preservation. The SIM-o values remain relatively stable across different severity levels, indicating that the synthesized speech maintains speaker identity. However, the autoPCP scores show a decreasing trend for more severe dysarthric speakers, implying that the prosody of synthesized speech deviates from the original as dysarthric severity increases. Overall, these results suggest that while the zero-shot TTS system can reasonably maintain speaker characteristics, it struggles to fully preserve the intelligibility and prosody of more severe dysarthric speech.

\vspace{-5pt}
\begin{figure}[b!]
    \centering
    \begin{subfigure}{}
        \centering
        \includegraphics[width=0.98\linewidth]{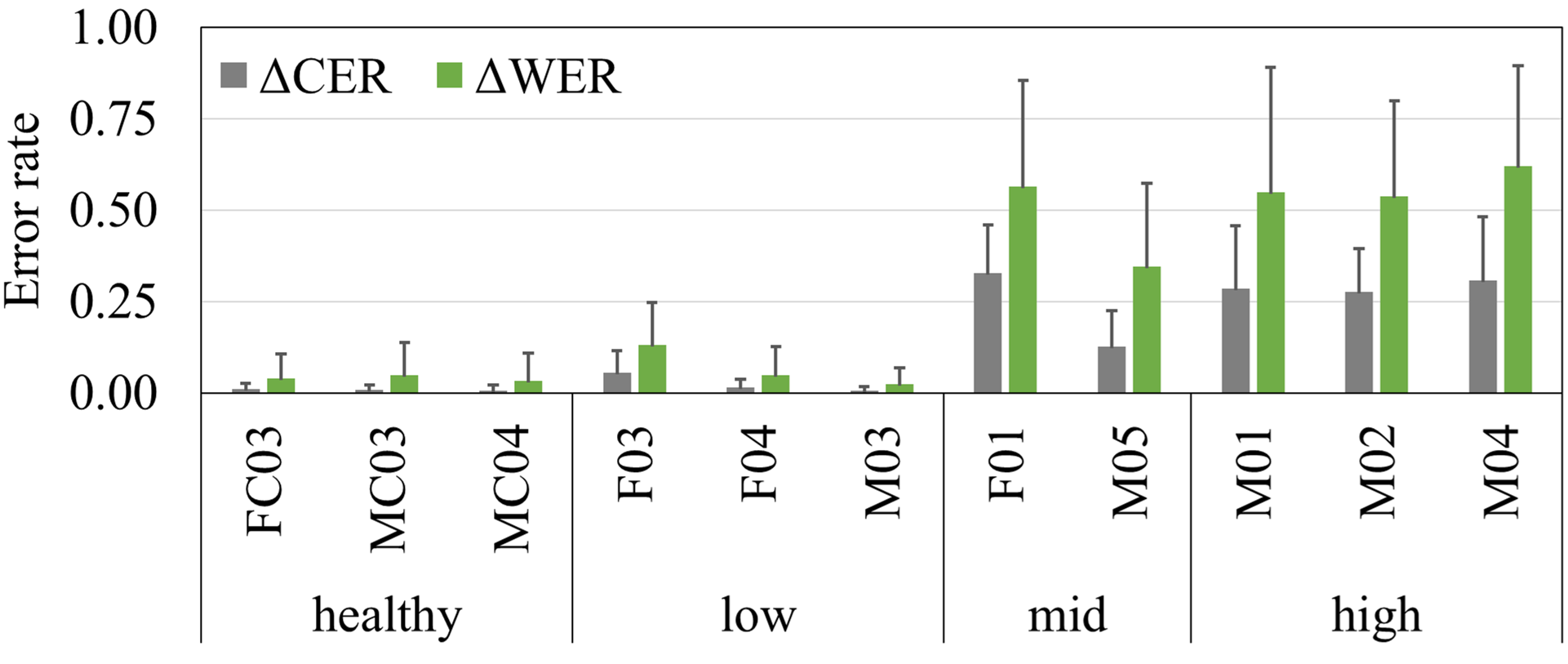}
        \footnotesize \text{ (a)$\Delta \text{WER}$ and $\Delta \text{CER}$}
    \end{subfigure}
    \begin{subfigure}{}
    \vspace{-8pt}
        \centering
        \includegraphics[width=0.98\linewidth]{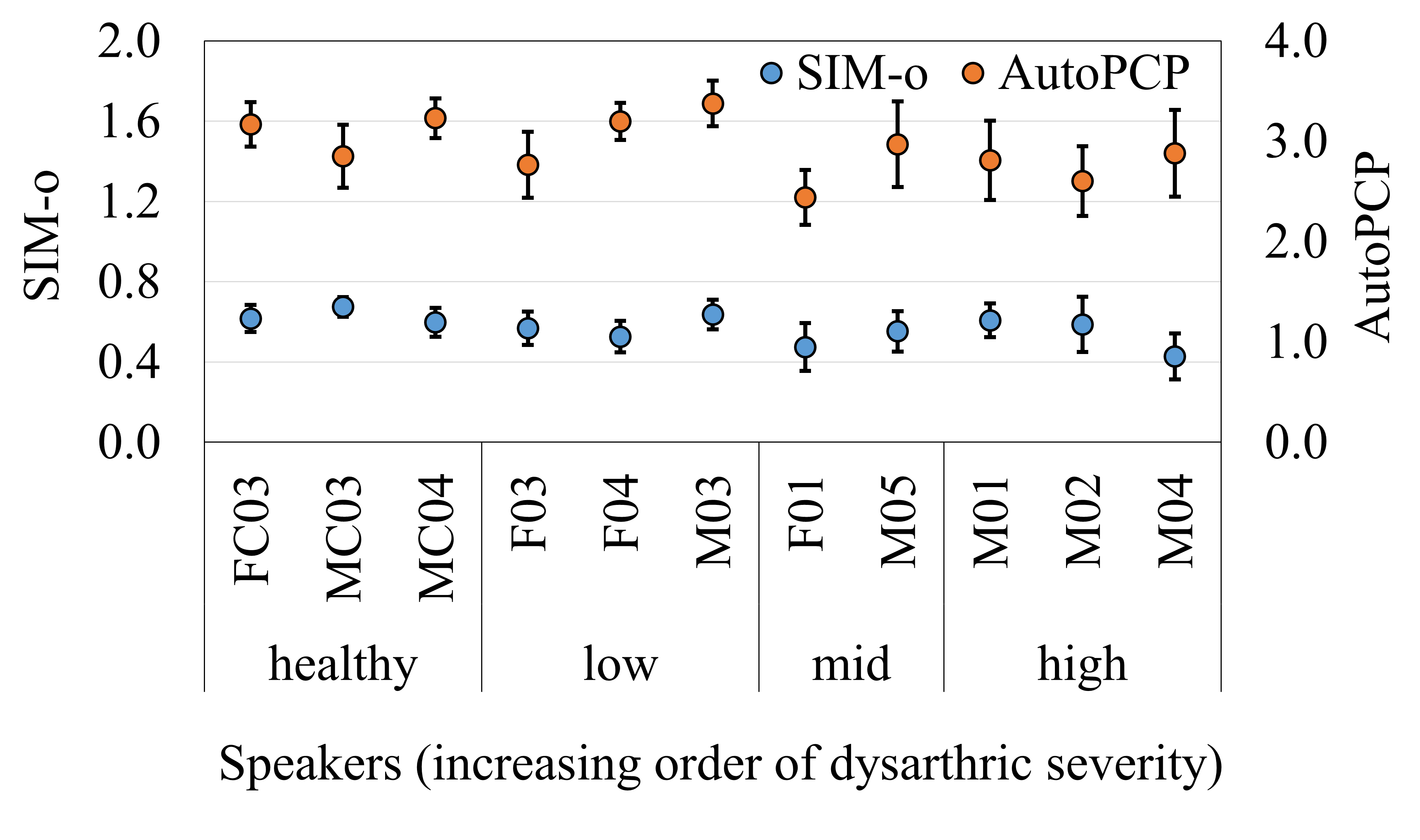}
        \footnotesize \text{(b) SIM-o and autoPCP}
        \vspace{-5pt}
    \end{subfigure}
    \caption{Objective assessment of zero-shot synthetic dysarthric speech using F5-TTS.}
    \vspace{-7pt}
\end{figure}
\vspace{-5pt}

\subsection{Fairness Metrics-Based Analysis}
In addition to our previous findings on the presence of biases in cloned dysarthric speech, this study employs fairness metrics to systematically analyze their impact across varying severity levels and gender categories, and the results are reported in Table~\ref{table:Fairness}. In general, speakers in the low severity category exhibit relatively clear speech patterns with only slight articulation difficulties, making their speech closely resemble that of healthy speakers. This is also reflected in the generated speech. As a result, minimal to no bias is being observed in both PD and DI across all objective measures. 

Apart from this, minimal bias is also observed in both the mid and high categories in retaining speaker and prosodic similarity, with PD remains closer to 0, and DI stays closer to 1 in both SIM-o and autoPCP. Conversely, high disparities are observed in $\Delta \text{WER}$  for both mid and high severity categories, where DI is less than 0.66 and PD is greater than 0.41. In contrast, for high severtiy category, while using $\Delta \text{CER}$, as it focuses on character-level articulation errors, a reduction of 27\% in DI and 85.71\% in PD is observed compared to $\Delta \text{WER}$. However, $\Delta \text{CER}$ still shows bias in the high severity category, with a DI of 0.75. Further, the fairness analysis across gender reveals a notable disparity, with male speakers exhibiting higher bias, particularly in intelligibility ($\Delta \text{WER}$) while female speakers shows higher bias in prosody (PD=0.36).

\begin{table}[h!]
\centering
\setlength{\tabcolsep}{3pt}
\caption{Fairness metrics based analysis for dysarthric severity and gender categories. \blue{High bias} is highlighted.}
\label{table:Fairness}
\vspace{-7pt}
\begin{tabular}{ccccccccc}
\hline
\multicolumn{9}{c}{Fairness across severity levels} \\ \hline
\multicolumn{1}{c|}{\multirow{2}{*}{\begin{tabular}[c]{@{}c@{}}Severity\\ category\end{tabular}}}                                        & \multicolumn{2}{c|}{$\Delta \text{WER}$}                            & \multicolumn{2}{c|}{$\Delta \text{CER}$}                            & \multicolumn{2}{c|}{SIM-o}                          & \multicolumn{2}{c}{autoPCP}    \\ \cline{2-9} 
\multicolumn{1}{c|}{}                                                                 & \multicolumn{1}{c|}{PD} & \multicolumn{1}{c|}{DI}   & \multicolumn{1}{c|}{PD} & \multicolumn{1}{c|}{DI}   & \multicolumn{1}{c|}{PD} & \multicolumn{1}{c|}{DI}   & \multicolumn{1}{c|}{PD} & DI   \\ \hline
\multicolumn{1}{c|}{healthy}                                                          & 0.0                     & \multicolumn{1}{c|}{1.0}  & 0.0                     & \multicolumn{1}{c|}{1.0}  & 0.0                     & \multicolumn{1}{c|}{1.0}  & 0.0                     & 1.0  \\
\multicolumn{1}{c|}{low}                                                         & 0.02                    & \multicolumn{1}{c|}{0.97} & 0.01                    & \multicolumn{1}{c|}{0.98} & 0.05                    & \multicolumn{1}{c|}{0.91} & 0.0                     & 0.99 \\
\multicolumn{1}{c|}{mid}                                                              & \blue{0.41}                    & \multicolumn{1}{c|}{\blue{0.66}} & \blue{0.22}                    & \multicolumn{1}{c|}{0.80} & 0.11                    & \multicolumn{1}{c|}{0.81} & 0.09                    & 0.87 \\
\multicolumn{1}{c|}{high}                                                              & \blue{0.52}                   & \multicolumn{1}{c|}{\blue{0.59}} & \blue{0.28}                    & \multicolumn{1}{c|}{\blue{0.75}} & 0.09                    & \multicolumn{1}{c|}{0.85} & 0.08                    & 0.90 \\ \hline 
\multicolumn{9}{c}{Fairness across gender}                                                                                                                                                                                                                              \\ \hline
\multicolumn{1}{c|}{\begin{tabular}[c]{@{}c@{}}Male\end{tabular}}   & \blue{0.37}                    & \multicolumn{1}{c|}{\blue{0.68}} & 0.19                    & \multicolumn{1}{c|}{0.82} & 0.07                    & \multicolumn{1}{c|}{0.88} & 0.11                    & 0.96 \\ 
\multicolumn{1}{c|}{\begin{tabular}[c]{@{}c@{}}Female\end{tabular}} & 0.21                    & \multicolumn{1}{c|}{0.81} & 0.12                    & \multicolumn{1}{c|}{0.88} & 0.09                    & \multicolumn{1}{c|}{0.84} & \blue{0.36}                    & 0.88 \\ \hline
\end{tabular}
\vspace{-15pt}
\end{table}

\subsection{Impact of Zero-Shot Voice-Cloning Based Augmentation on Downstream Tasks}

To understand the impact of zero-shot TTS augmented data with implicit biases, we have analysed its effect on the performance of downstream tasks such as ASR and dysarthria detection, and the results are presented in Table~\ref{tabel:tasks}. 
Notably, model~3 shows improvement, benefiting from the augmented data when trained alongside the reference audio samples. This results in an 18\% improvement in overall detection accuracy and a 44.57\% reduction in WER for the low severity category, compared to model~1 (model without augmentation). Conversely, due to the high bias towards higher severities, the addition of synthetic samples does not result in further improvement in the ASR task, instead leading to a reduction of 5.74\% and 7.65\% in WER for mid and high severities, respectively. As for model~2 being trained solely on synthetic data, it fails to generalise on the reference audio of dysarthric speech~\cite{hilmes24_syndata4genai}, leading to a degradation in performance. Overall, the addition of synthetic data benefits the dysarthria detection task more than the ASR task. Further research is required to better understand the effects.

\begin{table}[h!]
\centering
\small\setlength{\tabcolsep}{3pt}
\caption{Performance on downstream tasks. (model~1-trained on reference audio from TORGO, model~2-trained on synthetic audio, model~3-trained on both reference audio and synthetic audio.) \blue{Improvement} is highlighted.}
\label{tabel:tasks}
\vspace{-7pt}
\hbadness=10000
\hfuzz=6pt
\begin{tabular}{c|ccc|c}
\hline
     & \multicolumn{3}{c|}{Task1: ASR (WER in \%) } & \multirow{2}{*}{\begin{tabular}[c]{@{}c@{}}Task2: Dysarthria \\detection (Acc. in \%)\end{tabular}} \\ \cline{1-4}
Experiments & low      &    mid      & high    &                                                                                \\ \hline
model~1   & 53.00    & 89.07    & 87.53   & 62.50                                                                          \\
model~2   & 76.69    & 94.11    & 96.62   & 48.75                                                                          \\
model~3   & \blue{36.66}    & 94.19    & 94.23   & \blue{73.75}                                   
\\ \hline

\end{tabular}
\vspace{-15pt}
\end{table}

\subsection{Observations based on Fairness Study} 

Based on the fairness analysis and the observations from downstream tasks, the key insights are presented in Table~\ref{table:key_insights}. As discussed in the above sections, the results confirm that speech intelligibility (in terms of WER and CER) is the most biased aspect of dysarthric speech synthesis, particularly for severe cases. Unlike intelligibility, speaker and prosody similarity remain relatively unbiased, with DI above 0.81. These findings emphasize the need for fairness-aware data augmentation when using zero-shot voice cloning systems to generate synthetic speech for severe dysarthric speakers.


\section{Conclusion}
In this work, we presented a framework for evaluating biases in synthetic speech generated by F5-TTS for dysarthric speakers. Using fairness metrics, our analysis provides valuable insights into how dysarthric severity affects the preservation of intelligibility, speaker identity, and prosody in voice-cloned samples. While speaker similarity and prosodic information are retained across all severity levels, intelligibility is compromised in high severity cases with PD of 0.52 and DI of 0.59. This in turn affects the representation of higher dysarthric severities in data augmentation. Moreover, fine-tuning ASR models with such augmented data leads to performance degradation, particularly for individuals with more severe dysarthria. These findings strongly suggest the critical need for fairness-aware approaches in dysarthric speech augmentation. Future work will focus on exploring such methodologies to enhance the effectiveness of speech technologies for individuals with dysarthria.



\newpage
\bibliographystyle{IEEEtran}
\bibliography{refs}

\end{document}